\documentclass[aps,pra,reprint,superscriptaddress]{revtex4-1}

\usepackage{amsmath}
\usepackage{amssymb}
\usepackage{graphicx}
\usepackage{hyperref}
\usepackage{blindtext}

\begin{abstract}
We present a detailed study of the fields and propagation characteristics around the focus of ultrashort radially-polarized laser beams (RPLBs) having low-order spatio-temporal couplings (STCs). The three STCs considered are the focusing of the different frequencies to different positions along the longitudinal coordinate, the focusing of the frequencies to different positions along one transverse coordinate, and the beam waist or Rayleigh range having a dependence on frequency. The STCs considered are deemed low-order because they are primarily linear in frequency. The combination of a low-order vector beam, ultrashort pulse duration, and the three different STCs shows promise for exotic applications in dielectric or charged particle manipulation and potentially other strong-field phenomena. The STCs presented are all developed in a standard frequency-domain model where each case involves a different chromatic term. We present the results unique to the vector nature of the RPLBs and compare them to the linearly polarized cases, opening up opportunities for control of the electric field around the focus with the additional element of polarization.
\end{abstract}
\begin{document}

\title{Focused fields of ultrashort radially-polarized laser pulses having low-order spatio-temporal couplings}
\author{Spencer W. Jolly}
\email{spencer.jolly@vub.be}
\affiliation{LIDYL, CEA, CNRS, Universit{\'e} Paris-Saclay, CEA Saclay, 91191 Gif-sur-Yvette, France}
\affiliation{Brussels Photonics Team (B-PHOT), Dept. of Applied Physics and Photonics, Vrije Universiteit Brussel, Pleinlaan 2, 1050 Brussels, Belgium}
\date{\today}
\maketitle

\section{Introduction}
\label{sec:intro}

Radially-polarized laser beams (RPLBs), a unique solution to Maxwell's equations with primarily radial polarization, are a low-order vector beam (having spatially-varying polarization) that can have a very tight focus~\cite{quabis00,dorn03} and have applications in microscopy~\cite{sheppard04,lu09,kozawa11} and manipulation of micro-particles~\cite{zhanQ04,kawauchi07,kozawa10,huangL12} for example. Ultrashort RPLBs require additional modeling due to their broad bandwidth and can be applied to particle acceleration~\cite{varin05,fortin10,wong10,payeur12,carbajo16,wong17-3} and potentially high-harmonic generation and other realms of high-field nano-optics~\cite{dombi20}. Combing these ultrashort RPLBs with spatio-temporal couplings (STCs)---aberrations that lead to unseparable space-time or space-frequency electric fields~\cite{akturk10}---results in interesting properties. In this manuscript we present frequency-domain models to describe RPLBs having different STCs and investigate potentially valuable phenomena unique to the vector nature of RPLBs.

We first describe the standard ultrashort RPLB in frequency space, and describe it's basic properties. Then we compare certain properties of the RPLB when it has three different low-order STCs (i.e. generally linear in frequency dependence). The three STCs considered are: the focusing of the different frequencies to different positions along the longitudinal coordinate, longitudinal chromatism; the focusing of the frequencies to different positions along one transverse coordinate, spatial chirp; and the beam waist or Rayleigh range having a power-law frequency dependence; referred to as frequency-dependent beam parameters.

In the case of each STC there are general effects around the focus, which often involve an increase in the integrated beam size or an increase in the pulse duration (and equivalently a decrease in the intensity), similar to with linearly-polarized pulses~\cite{bourassin-bouchet11}. But in addition to those general effects there are more nuanced effects in each case, which can involve the pulse temporal profile, spatial intensity profile, or it's behavior during propagation away from the focus. In the case of longitudinal chromatism there is an effective increase in the Rayleigh range due to the longitudinal separation of the frequencies, and an additional effect when combined with temporal chirp where the velocity of the intensity peak can be different than the speed of light $c$, called the flying focus. For spatial chirp there can be wavefront rotation, where the wavefront in focus points in different directions over time, and there can also be a tilt in the arrival time of the pulse when combined with temporal chirp. In the case of the frequency-varying beam parameters we consider the evolution of the carrier-envelope offset phase, which differs from the Gouy phase in a non-trivial fashion.

Indeed all of the above phenomena have been described and measured experimentally for linearly-polarized ultrashort laser pulses. In the case of the RPLBs which we describe theoretically here, there is first the fact that the model for the fields is different, resulting in different precise descriptions for each phenomenon. But there is also the crucial fact that RPLBs are vector beams and the different polarization components of the electric field not only have different basic properties that influence how to look at the effect of STCs on each polarization, but the different components are also described via different equations resulting in different behavior. In each case we develop the model for the RPLB with STCs using the same frequency-space model as for the standard ultrashort RPLB, but clearly explain the modified or added terms which result in the fundamental changes in beam properties or propagation effects. Because the STCs add non-trivial space-frequency terms in our model we do not have the ability to solve for the fields in time, but can present results in time based on our model using numerical Fourier-transforms.

\section{Standard ultrashort RPLB}
\label{sec:RPLB}

In the following model we use pulses that have Gaussian spatial and spectral/temporal profiles, with characteristic widths $s_0$ and $\tau_0$ respectively, at a central wavelength $\lambda_0$ ($\omega_0=2\pi c/\lambda_0$) and the Rayleigh range $z_R=\omega_0 s_0^2/2c$. The fields of the focused ultrashort RPLBs are modeled in the frequency domain as in Refs.~\cite{sainte-marie17,jolly19-1,jolly20-2} using the proper form for the longitudinal fields including only paraxial terms. With $A_\omega=\exp(-\delta\omega^2/\Delta\omega^2)$, $\Delta\omega=2/\tau_0$, and $\delta\omega=(\omega-\omega_0)$ we have the solutions for the radial electric field $E_r$, longitudinal electric field $E_z$, and azimuthal magnetic field $B_{\theta}$

\begin{align}
\begin{split}
&\tilde{E}_r(\omega)=A\epsilon\rho C_{2} \label{eq:E_r}
\end{split}\\
\begin{split}
&\tilde{E}_z(\omega)=A\epsilon^{2}\left[S_{2}-\rho^{2} S_{3}\right] \label{eq:E_z}
\end{split}\\
\begin{split}
&\tilde{B}_{\theta}(\omega)=\frac{\tilde{E}_r(\omega)}{c} \label{eq:B_t},
\end{split}
\end{align}

\noindent where the tilde denotes frequency space, $\rho=r/s_0$ and $\epsilon=s_0/z_R$, and

\begin{align}
A&=\frac{\omega_0}{2c}\sqrt{\frac{8P_0}{\pi \epsilon_0 c}} \frac{\sqrt{2}A_\omega}{\Delta\omega} e^{-{r}^2/s^2} \label{eq:A} \\
C_n&=\left(\frac{s_0}{s}\right)^n e^{i(\psi + n\psi_G)} \label{eq:C_n}\\
S_n&=\left(\frac{s_0}{s}\right)^n e^{i(\psi + n\psi_G + \pi/2)} \label{eq:S_n}\\
\psi&=\Psi_0-\frac{\omega {r}^2}{2cR}-\frac{\omega{z}}{c}-\psi_\textrm{spec} \label{eq:psi}\\
\psi_\textrm{spec}&=\frac{\phi_2\delta\omega^2}{2}+\frac{\phi_3\delta\omega^3}{6}+\frac{\phi_4\delta\omega^4}{24}+... \label{eq:psi_spec}\\
\psi_G&=\tan^{-1}\left(\frac{z}{z_R}\right) \label{eq:psi_G}\\
s&=s_0\sqrt{1+\left(\frac{z}{z_R}\right)^2} \label{eq:w}\\
R&=z+\frac{z_R^2}{z} \label{eq:R}.
\end{align}

\noindent $P_0$ is the peak power of the pulse (in this case with zero STCs), $\epsilon_0$ is the permittivity of free space, and $c$ is the speed of light in vacuum, however for the rest of this work we will present normalized fields since we are modeling their properties in free space only. The spectral phase parameters $\phi_2$, $\phi_3$, and $\phi_4$ are the group-delay dispersion, third-order dispersion, and fourth-order dispersion respectively, included for completeness, but in our presented phenomena we will only show results with nonzero $\phi_2$. The fields above are the complex fields, where for all visualizations we show the real part of the fields in time corresponding to the physically relevant quantities.

Of course in the case of the equations above, the fields in time can be easily calculated. This is because the only frequency dependence is in the Gaussian spectral profile $A_{\omega}$ and the term $\omega z/c$ in the phase, so the temporal profile is simply a Gaussian envelope traveling in $z$ at the speed of light. However, for the rest of this work we will build off of this standard case to consider pulses with non-trivial chromatic terms, which cannot be expressed simply in time, so we leave these fields in terms of frequency for reference. Because it is often the electric fields that are most important, $B_{\theta}$ will not generally be included in visualizations, except for in the case of spatial chirp where calculating the magnetic field requires a different coordinate transformation due to the breaking of cylindrical symmetry. It must also be noted that modeling an azimuthally-polarized laser beam requires only a simple transformation of these equations, so the work in the rest of this manuscript will also apply.

\begin{figure}[tb]
	\centering
	\includegraphics[width=86mm]{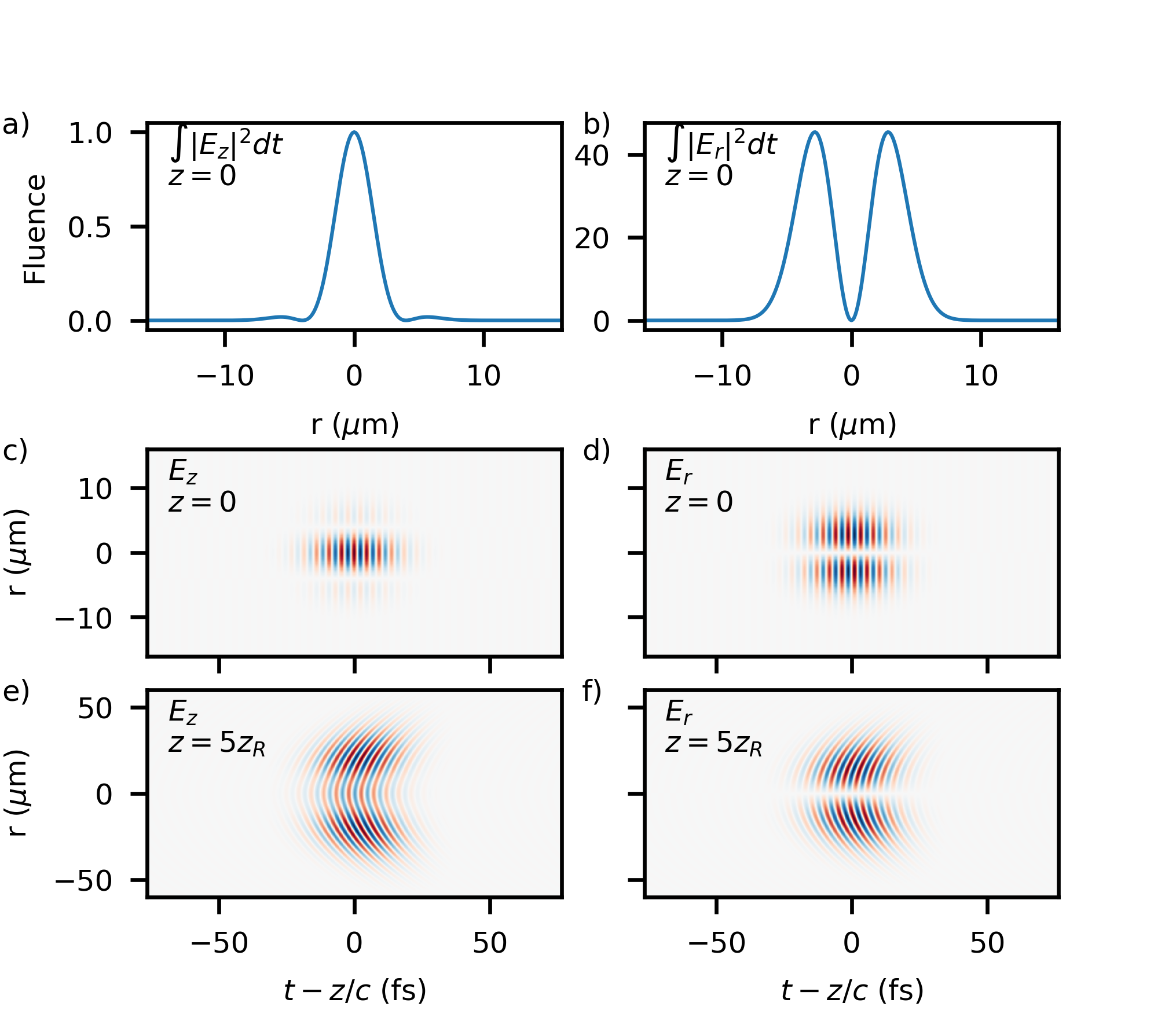}
	\caption{Standard ultrashort RPLB in-focus with $\lambda_0=800$\,nm, $\tau_0=15$\,fs, and $s_0=4$\,$\mu$m. The integrated $E_z$ (a) is strongly localized on-axis, where the integrated $E_r$ (b) has a zero on-axis. The $E_z$ and $E_r$ fields in time, (c) and (d) respectively, show similar characteristics, but as they diffract they change --- shown at $z=5z_R$ in (e) and (f) respectively. The fluence in (a) and (b) is normalized to the max fluence of $E_z$, and the color scale in each of panels (c)--(f) is relative to the maximum in that panel.}
	\label{fig:RPLB_standard}
\end{figure}

Figure~\ref{fig:RPLB_standard} shows the basic characteristics of the standard ultrashort RPLB for a wavelength of 800\,nm that we will use throughout this manuscript. These characteristics include the longitudinal field $E_z$ being strongly localized on-axis and the transverse field $E_r$ being zero on-axis, shown in Fig.~\ref{fig:RPLB_standard}(a) and (b) respectively, with both having cylindrical symmetry. As already mentioned, since this description does not yet contain STCs the fields are trivially extended to having an ultrashort envelope that maintains these features in the focus, shown in Fig.~\ref{fig:RPLB_standard}(c)--(d). Both the monochromatic RPLB and the ultrashort RPLB become more complicated as they propagate away from the focus, shown for the ultrashort case in Fig.~\ref{fig:RPLB_standard}(e)--(f), still without STCs.

Note as well that the non-paraxial form could be easily constructed by adding terms of higher-order in $\epsilon$ as in Ref.~\cite{salamin06} for example. This will also be the case for all of the models used in the following sections with STCs added, but we write only the paraxial forms for simplicity in the main text to emphasize the chromatic terms. For a more detailed discussion of the non-paraxial description see the appendix. An accurate non-paraxial description is generally believed to be necessary when the beam waist approaches the wavelength, but for highly sensitive interactions such as high-field particle acceleration a non-paraxial description becomes important even at larger beam waists~\cite{marceau13-1}.

\section{Longitudinal Chromatism}
\label{sec:LC}

Longitudinal chromatism (LC) is a form of spatio-temporal coupling that is when the different spectral components of an ultrashort pulse are separated longitudinally, that is, along the direction of propagation $z$. In the case in this section, we model LC at the focus of an ultrashort RPLB. In fact, LC is the result of two equivalent phenomenon on the collimated beam before focusing, pulse-front curvature (PFC) and chromatic curvature (CC). PFC is when the arrival time of the ultrashort pulse varies quadratically with the radius. Due to the fundamental relationship between time and frequency in light waves (via the Fourier transform), PFC is equivalent to the radius of curvature of the pulse depending linearly on frequency, which we refer to as chromatic curvature (CC). Because of this equivalence, we refer to this phenomenon on the collimated beam as CC/PFC, which is denoted by $\alpha$~[$fs/m^2$]. 

\begin{figure}[tb]
	\centering
	\includegraphics[width=86mm]{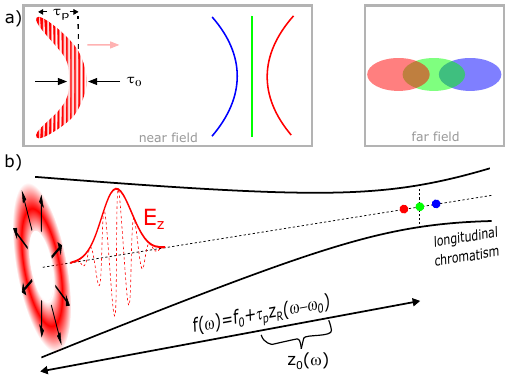}
	\caption{Basic concept of longitudinal chromatism (LC) in the focus. A beam in the near field (a) before being focused can have pulse-front curvature (PFC) where the arrival time depends quadratically on the radius, and equivalently the frequencies have a linearly varying convergence. When focused the frequencies focus to different longitudinal positions and produce LC, shown on the right in (a). The scenario relevant for modeling the RPLB is shown schematically in (b), now with multiple components of the field.}
	\label{fig:LC_concept}
\end{figure}

Because the manifestation of the near-field couplings in focus depends on the focusing geometry, the corresponding near-field beam radius can be chosen as $s_i$ with a focal length of $f$, making $s_0=2cf/\omega_0 s_i$. However, if the collimated beam were smaller and the focal length shorter so as to have the same beam waist and Rayleigh range, then the relative effect of a certain CC/PFC $\alpha$ would be different. So in fact, it is the quantity $\tau_p=\alpha s_i^2$ (see Fig.~\ref{fig:LC_concept}(a)) that can properly parameterize the effect of the CC/PFC coupling regardless of the actual focusing geometry. So it is this $\tau_p$ that we use to parameterize the LC in-focus (resulting from CC/PFC on the collimated beam) for the rest of this section. The LC produces a frequency-dependent longitudinal waist position $z_0$ of

\begin{equation}
z_0=z_R\tau_p\delta\omega,
\end{equation}

\noindent where $\delta\omega=(\omega-\omega_0)$, i.e. describing linear LC.

The fields around the focus $z=0$ are as before in Eqs.(\ref{eq:E_r})--(\ref{eq:psi_spec}), but with modified terms

\begin{align}
	\psi_G&=\tan^{-1}\left(\frac{z-z_0}{z_R}\right) \label{eq:psi_G_LC}\\
	s&=s_0\sqrt{1+\left(\frac{z-z_0}{z_R}\right)^2} \label{eq:w_LC}\\
	R&=\left(z-z_0\right)+\frac{z_R^2}{\left(z-z_0\right)} \label{eq:R_LC}.
\end{align}

We must note that the waist $s_0$ and Rayleigh range $z_R$ being constant values relies on the pulses being longer than few-cycle as mentioned earlier, but in the case of LC it also relies on the difference in focal lengths between the extreme frequencies of the pulse being negligible. This is equivalent to the assumption that the extended Rayleigh range ${z_R}^e=\tau_p\Delta\omega{z_R}$ is negligible compared to the focal length $f$: $\tau_p\Delta\omega{z_R}\ll{f}$ or $\tau_p/\tau_0\ll{f}/{2z_R}$. If this were not the case, then the focal length would also need to be treated as dependent on frequency, resulting in a frequency dependence of the waist and Rayleigh range independent of the assumption of the pulse duration. After calculating the fields in frequency space with the stated assumptions, they must be inverse Fourier transformed to time.

\begin{figure}[tb]
	\centering
	\includegraphics[width=86mm]{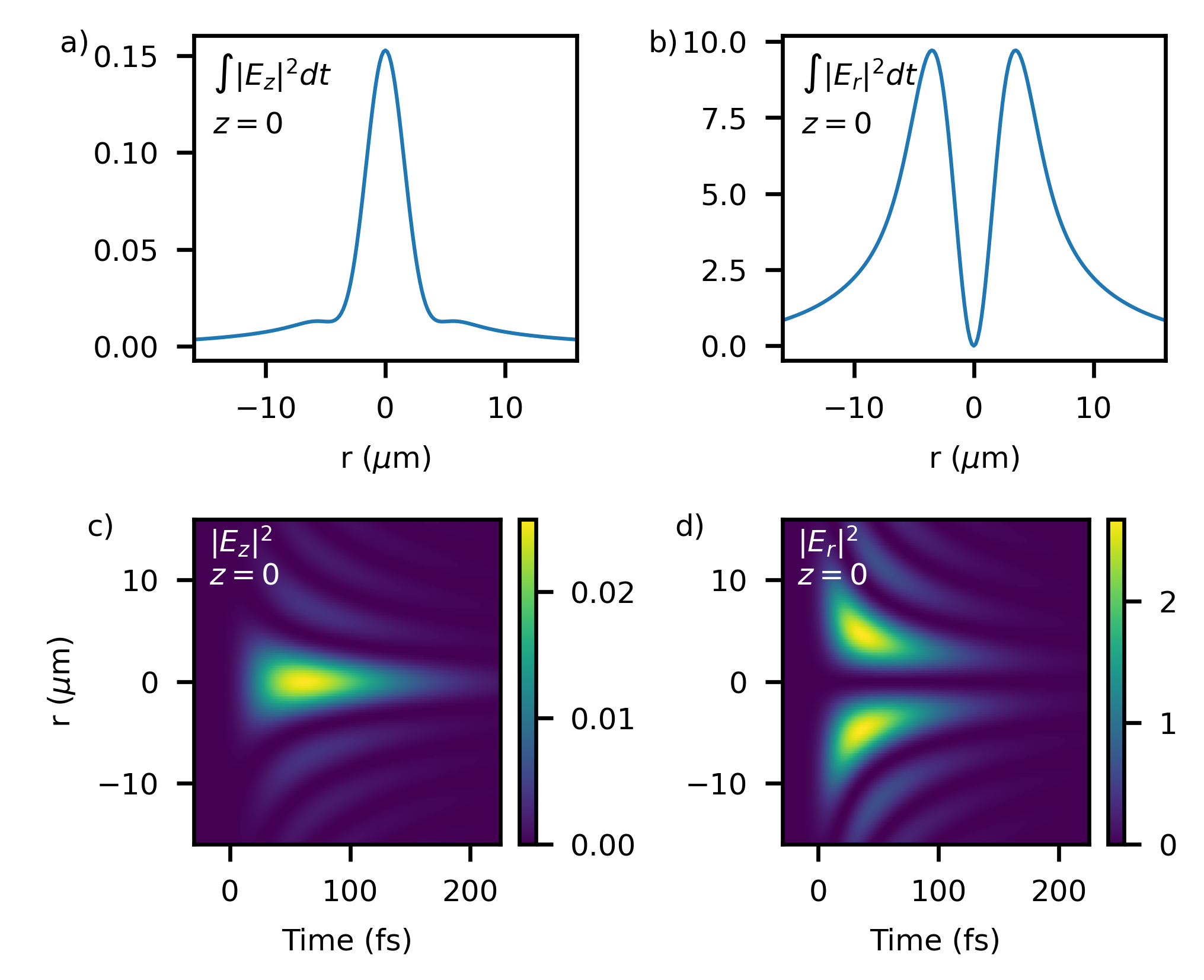}
	\caption{Basic results with longitudinal chromatism $\tau_p=60$\,fs ($\lambda_0=800$\,nm, $\tau_0=15$\,fs, $s_0=4$\,$\mu$m). The integrated $E_z$ and $E_r$ fields in-focus, (a) and (b) respectively, show larger tails as $r$ increases. The $E_z$ and $E_r$ fields in time, (c) and (d) respectively, now have a time asymmetry and fringes at larger $r$ characteristic of chromaticity, but the qualitative characteristics stay the same. The fluence in (a) and (b) is normalized to the max fluence of $E_z$ without any LC and the intensity in (c) and (d) is normalized to the max intensity of $E_z$ without any LC.}
	\label{fig:LC_results}
\end{figure}

The basic results of the model of an ultrashort RPLB with LC is shown in Figure~\ref{fig:LC_results} for one value of $\tau_p=60$\,fs in the focus at $z=0$. The pulse retains cylindrical symmetry, but both $E_z$ and $E_r$ have much longer-tailed transverse distributions, shown in Fig.~\ref{fig:LC_results}(a)--(b). This is seen very clearly when comparing Fig.~\ref{fig:LC_results}(a)--(b) directly to Fig.~\ref{fig:RPLB_standard}(a)--(b), where the only difference is the LC.

Besides the effect on the transverse profile of both polarization components, there is also a significant change in the temporal profile. As shown in Fig.~\ref{fig:LC_results}(c)--(d) the temporal intensity profiles of both $E_z$ and $E_r$ are no longer symmetric in time, and have significant interference effects at the outer portions of the transverse distributions. The asymmetry and interferences are common characteristics of LC, and in this case are manifested in both polarization components, while each component maintains the basic characteristics of the ultrashort RPLB --- $E_z$ is localized on-axis and $E_r$ is zero on-axis.

Another feature of the ultrashort RPLB with LC is an extended Rayleigh range, defined earlier as ${z_R}^e=\tau_p\Delta\omega{z_R}$. This can be simply reasoned from the fact that different colors have their waist position at different $z$ positions. So, although the waist is larger due to the LC, the pulse retains that slightly larger waist within an increased longitudinal distance ${z_R}^e$, larger than the basic Rayleigh range by both the amount of LC $\tau_p$ and the bandwidth of the pulse $\Delta\omega$.

\begin{figure}[tb]
	\centering
	\includegraphics[width=86mm]{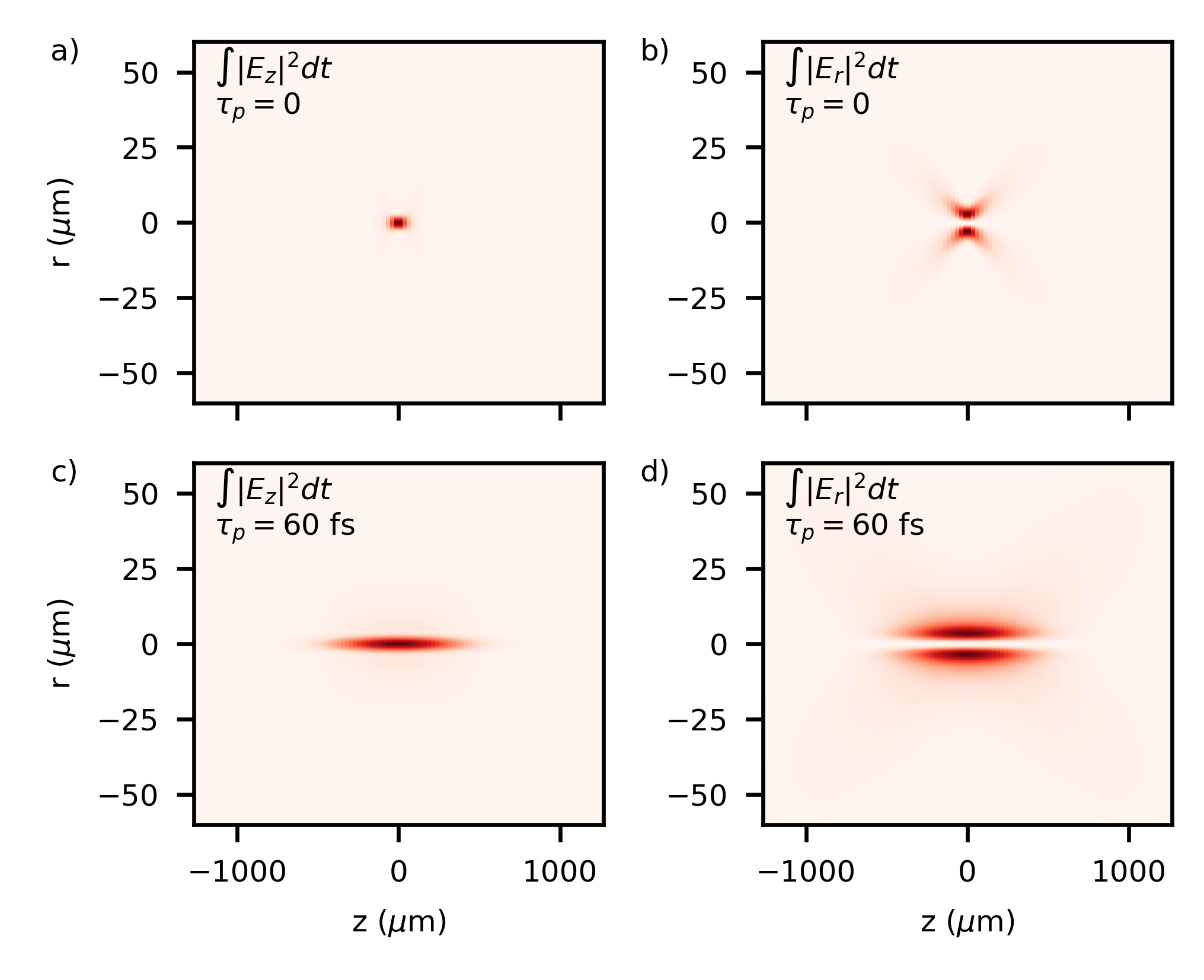}
	\caption{Increased Rayleigh range due to the longitudinal chromatism of both the longitudinal field $E_z$, (a) with no LC and (c) with $\tau_p=60$\,fs, and transverse field $E_r$, (c) with no LC and (d) with $\tau_p=60$\,fs ($\tau_0=15$\,fs, $s_0=4$\,$\mu$m).}
	\label{fig:LC_zRe}
\end{figure}

This increased Rayleigh range, effectively beating diffraction, has already been shown for linearly polarized pulses~\cite{froula18}. We confirm in Figure~\ref{fig:LC_zRe} that it is also valid for ultrashort RPLBs having LC for both $E_z$ and $E_r$. Comparing Fig.~\ref{fig:LC_zRe}(a)--(b) without LC to Fig.~\ref{fig:LC_zRe}(c)--(d) with $\tau_p=60$\,fs the Rayleigh range is roughly $2\tau_p/\tau_0=8$ times larger in the latter case. What is not shown in Fig.~\ref{fig:LC_zRe}(c)--(d) of course is the temporal behavior of the electric field at each $z$-position, as it is only showing the integrated intensity. Because the LC separates the colors then at the different points within the extended Rayleigh range the central wavelength will evolve, which can be thought of as the price paid for "beating" diffraction. Two-dimensional STC wavepackets~\cite{kondakci16,kondakci17} have been shown to remain localized for orders of magnitude beyond the diffraction limit without separating the frequencies~\cite{bhaduri18,bhaduri19-2}, but have other limitations and are not vector beams, so can be considered a uniquely different phenomenon.

\subsection{Flying focus effect}
\label{sec:LC_FF}

Longitudinal chromatism combined with linear temporal chirp (or quadratic spectral phase $\phi_2$) results in the intensity of the ultrashort laser pulse traveling at velocities significantly different than $c$, referred to as the flying focus~\cite{sainte-marie17,palastro18}. This has been implemented with either diffractive optics~\cite{froula18} or specially produced lens doublets~\cite{jolly20-1} to produce the LC, and has also been used to create ionization waves in a plasma at such superluminal and even negative velocities~\cite{turnbull18-2,franke19}. Theoretical work shows potential applications in laser-plasma Raman amplification~\cite{turnbull18-1}, plasma-based photon acceleration~\cite{howard19}, and vacuum electron acceleration~\cite{ramsey20}. All of the past work has dealt with scalar electric fields, so we discuss and simulate the same effect with the fields of an ultrashort RPLB.

The velocity of the intensity peak in the flying focus scenario $v_{ff}$ has been shown to obey a simple formula~\cite{sainte-marie17,palastro18,jolly20-1}

\begin{equation}
	\frac{v_{ff}}{c} = \frac{1}{1+\frac{c\phi_2}{\tau_p z_R}} \label{eq:LC_FF} .
\end{equation}

\noindent This simple formula is derived based on only the varying central frequency along the longitudinal direction and concurrent varying arrival time for the different frequencies resulting from the spectral phase. We expect that it holds similarly for the RPLB having chirp and LC since the relevant phase terms are the same.

\begin{figure}[tb]
	\centering
	\includegraphics[width=86mm]{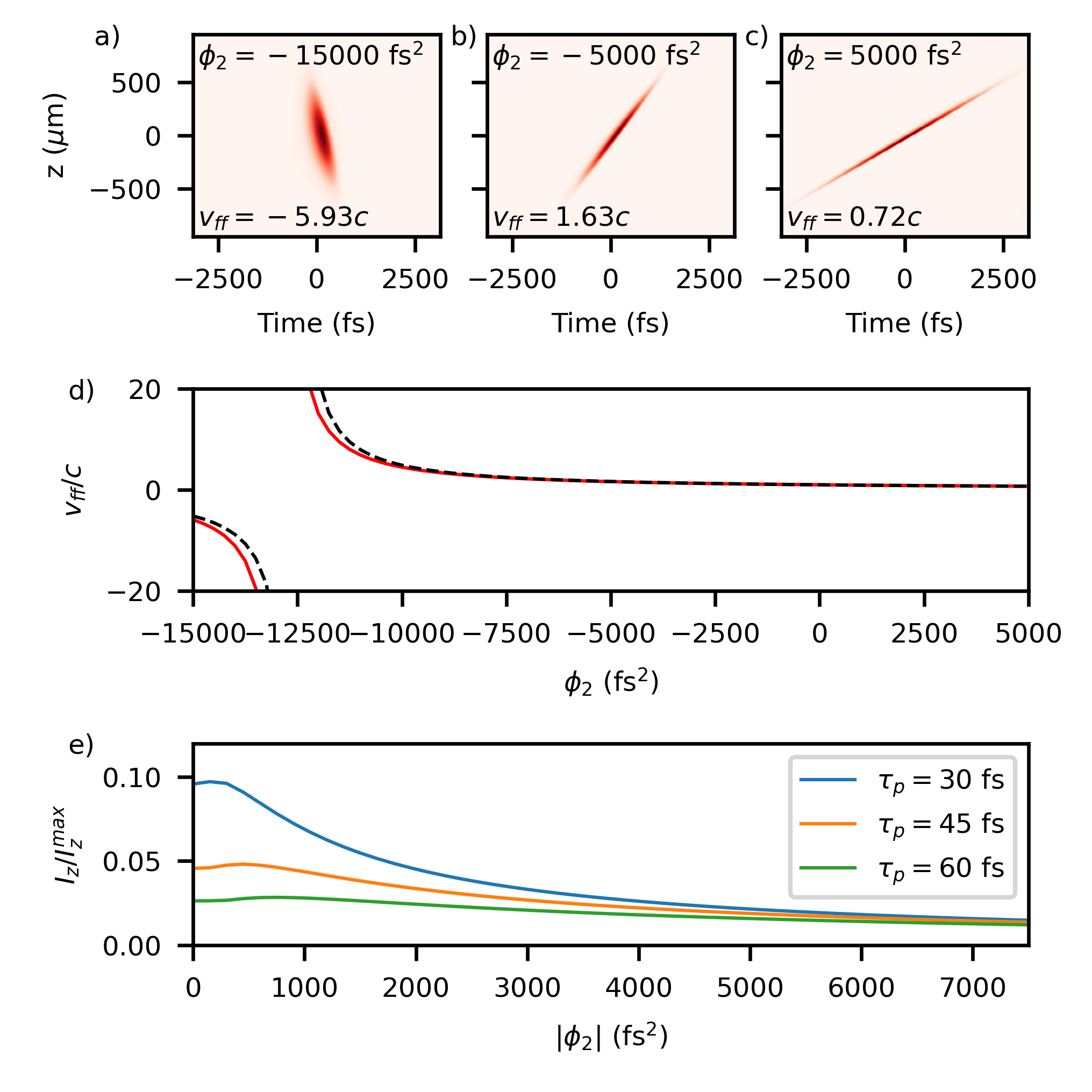}
	\caption{Flying focus effect with the longitudinal field $E_z$ at $r=0$. An RPLB with an LC of $\tau_p=60$\,fs ($\tau_0=15$\,fs, $s_0=4$\,$\mu$m) is simulated with varying linear chirp ($\phi_2$) to produce an intensity envelope that travels at velocities much different than $c$ (a)--(c). The dependence of the intensity peak velocity $v_{ff}$ on chirp (d) in the analytical equation Eq.~\ref{eq:LC_FF} (black dashed line) agrees with the full simulations (red solid line). The intensity decreases (e) from the case with no STCs with different values of $\tau_p$ and with increasing chirp.}
	\label{fig:LC_FF}
\end{figure}

The results for the combination of LC and chirp are shown in Figure~\ref{fig:LC_FF} for $E_z$ only, with $r=0$, confirming that the flying focus phenomenon occurs as well with ultrashort RPLBs. The maps of the on-axis intensity profile for $z$ positions through the focus is shown for $E_z$ with a single value of $\tau_p=60$\,fs and three chirp values in Fig.~\ref{fig:LC_FF}(a)--(c). These specific combinations of LC and chirp produce negative super-luminal, positive super-luminal, and positive sub-luminal $v_{ff}$ in Fig.~\ref{fig:LC_FF}(a), (b), and (c) respectively. The dependence of $v_{ff}$ on the chirp from the simulations agrees very well with that predicted from the simple relationship in Eq.~(\ref{eq:LC_FF}), shown in Fig.~\ref{fig:LC_FF}(d). The transition chirp where the velocity transitions from purely positive to negative occurs at roughly $\phi_2=-\tau_p z_R/c$, the same as for linear polarization. The peak intensity of the longitudinal field decreases significantly already with the LC, and decreases further with added chirp (Fig.~\ref{fig:LC_FF}(e)), which is qualitatively similar to the intensity reduction for linear polarization~\cite{sainte-marie17}.

\begin{figure}[tb]
	\centering
	\includegraphics[width=86mm]{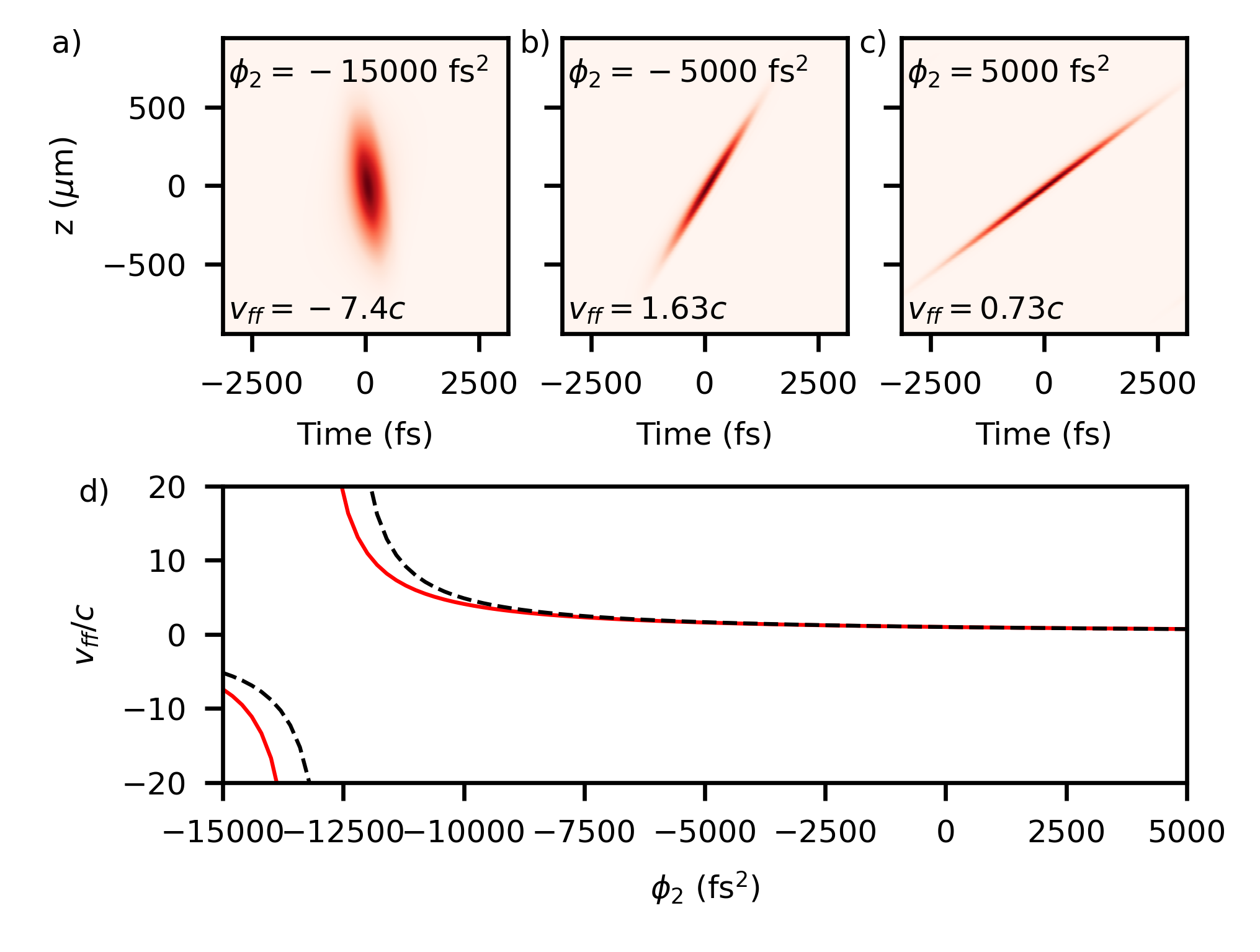}
	\caption{Flying focus effect with the transverse field $E_r$ at the off-axis position $r=s_0$. An RPLB with an LC of $\tau_p=60$\,fs ($\tau_0=15$\,fs, $s_0=4$\,$\mu$m) is simulated with varying linear chirp ($\phi_2$) to produce an intensity envelope that travels at velocities much different than $c$ (a)--(c), slightly different than those for $E_z$ on-axis. The dependence on $\phi_2$ (d) is slightly different in this case as well (red solid line), diverging more from the simple prediction (black dashed line) based on Eq.~\ref{eq:LC_FF}.}
	\label{fig:LC_FF_Er}
\end{figure}

The situation of LC and chirp in the transverse polarization, $E_r$, is more complicated both because the field is only non-zero off-axis, and because as the pulse diffracts away from the focus the position of maximum intensity is increasingly off-axis. However, as already shown in Fig.~\ref{fig:LC_zRe}(d), within the extended Rayleigh range the position of maximum intensity is relatively constant. Therefore we look at the velocity of the intensity peak of $E_r$ at $r=s_0$ ($\rho=1$). The results for the same nonzero chirps as for $E_z$ is shown for $E_r$ in Fig.~\ref{fig:LC_FF_Er}.

The result of the analysis of the flying focus off-axis with $E_r$ is that it is only slightly different than that of $E_z$, both in terms of the quantitative results and the qualitative dependence on $\phi_2$. This is due to the additional phase term dependent on $r$ in Eq.~(\ref{eq:psi}), the curvature term, that due to the transformation of $z$ to $z-z_0(\omega)$ introduces a non-trivial frequency dependence when $r\neq0$. When comparing the simulated temporal intensity profiles of $E_z$ at $r=0$ in Fig.~\ref{fig:LC_FF}(a)--(c) and for $E_r$ at $r=s_0$ in Fig.~\ref{fig:LC_FF_Er}(a)--(c), the results are therefore slightly different. However, this is most apparent around the transition chirp. For example with $\phi_2=-15000$\,fs$^2$ the velocity is -5.93$c$ for $E_z$ on-axis and -7.4$c$ for $E_r$ off-axis. The difference in the dependence on the chirp is seen clearly in Fig.~\ref{fig:LC_FF_Er}(d) near the transition chirp of -12500\,fs$^2$ when compared to Fig.~\ref{fig:LC_FF}(d). It is interesting that the phase term due to curvature is in the phase both for $E_z$ and $E_r$, meaning that as $r$ increases the velocity changes for $E_z$ as well, which we have confirmed via numerical calculations, but this difference is less important for $E_z$ because the intensity decreases rapidly off-axis.

As a last observation, both for $E_z$ and $E_r$, producing a changing velocity (i.e. accelerating, decelerating, or non-monotonic) of the intensity peak is also possible with higher orders of spectral phase $\phi_3$, $\phi_4$, etc. as already predicted for linear polarization~\cite{sainte-marie17}.

It is important again to emphasize the difference from STC wavepackets~\cite{kondakci17} that have been shown as well to have tunable velocities~\cite{kondakci19-2,bhaduri19-1,yessenov20-3}. The flying focus shown here, besides being with a cylindrical vector beam, is limited to a region around the focus defined by $z_{R}^e$ and the different frequencies of the initial laser pulse are separated longitudinally within that region. Therefore, the flying focus for both linear polarization and with ultrashort RPLBs is most relevant to phenomena that are purely intensity-dependent, otherwise this separation of the frequencies must be taken into account. The tunable velocity STC wavepackets, shown so far only for linear polarization, do not have the separation of frequencies and can have localized propagation orders of magnitude longer than as predicted by diffraction, but for reasons already discussed can be considered significantly different than the case discussed in this manuscript. Besides these STC wavepackets, more complex examples with programmable group velocities involving Bessel beams have recently been described~\cite{li20}, which could also theoretically be adapted to RPLBs.

\section{Spatial Chirp}
\label{sec:SC}

\begin{figure}[tb]
	\centering
	\includegraphics[width=86mm]{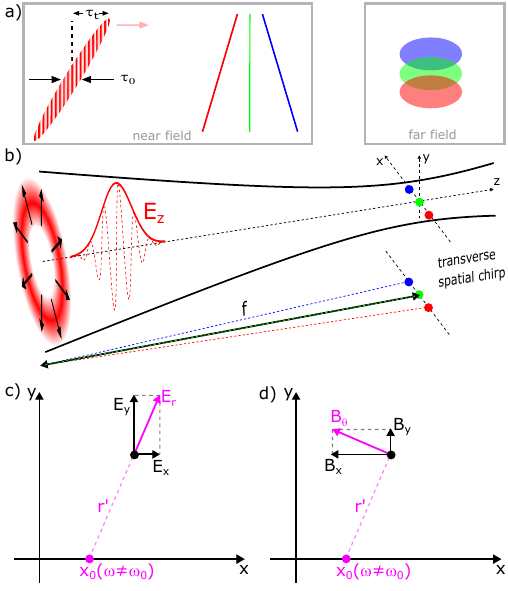}
	\caption{Basic concept of spatial chirp (SC) in the focus. A beam in the near field (a) before being focused can have pulse-front tilt (PFT) where the arrival time depends linearly on one spatial coordinate, and equivalently the frequencies have a linearly varying pointing direction, or angular dispersion (AD). When focused (b) the frequencies separate along the same axis as the AD/PFT and produce spatial chirp (SC). Due to the breaking of cylindrical symmetry both $E_r$ and $B_{\theta}$ require a frequency-dependent transformation to cartesian coordinates, shown in (c) and (d) respectively.}
	\label{fig:SC_concept}
\end{figure}

In this section we model spatial chirp in the focus of an RPLB. Spatial chirp around the focus is due to angular-dispersion (AD) in the near-field, where different frequencies have different pointing directions, or equivalently the pulse-front of the beam is tilted from the wavefront (pulse-front tilt, PFT), see Fig.~\ref{fig:SC_concept}(a). Because these are equivalent, it can be referred to as AD/PFT to be unambiguous. When focused the different pointing directions of different frequencies result in the frequencies being spatially separated in the focal region, as in Fig.~\ref{fig:SC_concept}(b), along the axis where there was AD/PFT in the near field. This transverse spatial separation of frequencies is exactly spatial chirp (SC).

As it was with LC, the in-focus manifestation of the near-field AD/PFT---represented by the tilt angle $\eta=dt/dx$ [fs/m]---depends on the focusing geometry. So, similar to the case of LC it is the time delay at the outer edge of the near field (with beam size $s_i$) due to AD/PFT $\tau_t=\eta s_i$ (see Fig.~\ref{fig:SC_concept}(a)) that can properly parameterize the effect of the near-field AD/PFT coupling regardless of the actual focusing geometry. So it is this $\tau_t$ that we use to parameterize the SC in-focus (resulting from AD/PFT on the collimated beam) for the rest of this section. Linear SC is described by a best-focus position $x_0$ of

\begin{equation}
x_0=s_0\tau_t\delta\omega/2,
\end{equation}

\noindent which is clearly linear in frequency and proportional to $\tau_t$ representing the AD/PFT.

The key difference between the modeling of SC is that there is no longer cylindrical symmetry as in the previous section. This is because with AD/PFT on the collimated beam, the different frequencies are focused to different positions along one transverse axis, chosen to be $x$ here (see Fig.~\ref{fig:SC_concept}(b)). This not only affects the spectral content at any given position, but it affects the direction of the fields, which depends non-trivially on frequency.

The implication is that the initial fields $\tilde{E}_r$, $\tilde{E}_z$, and $\tilde{B}_{\theta}$ must be transformed to Cartesian coordinates, and all fields $\tilde{E}_x$, $\tilde{E}_y$, $\tilde{E}_z$, $\tilde{B}_x$, and $\tilde{B}_y$ ($\tilde{B}_z=0$ still) must be calculated at all $x$, $y$, and $z$ positions. However, because the frequencies are shifted a different amount along $x$ in the focal plane, the coordinate transformation is also frequency dependent. This concept is sketched in Fig.~\ref{fig:SC_concept}(c) for $\tilde{E}_r$ and in Fig.~\ref{fig:SC_concept}(d) for $\tilde{B}_{\theta}$. The results are:

\begin{align}
\tilde{E}_x(\omega)&=\tilde{E}_r(\omega, r')\left(\frac{x-x_0}{r'}\right) \label{eq:Cart1}\\
\tilde{E}_y(\omega)&=\tilde{E}_r(\omega, r')\left(\frac{y}{r'}\right) \label{eq:Cart2}\\
\tilde{B}_x(\omega)&=\tilde{B}_{\theta}(\omega, r')\left(\frac{-y}{r'}\right) \label{eq:Cart3}\\
\tilde{B}_y(\omega)&=\tilde{B}_{\theta}(\omega, r')\left(\frac{x-x_0}{r'}\right) \label{eq:Cart4},
\end{align}

\noindent where $r'=\sqrt{(x-x_0)^2+y^2}$. The $\tilde{E}_z$ field does not require any transformation, but it is still distributed along $x$ according to $x_0$.

As before we use pulses that have Gaussian spatial and temporal profiles, with characteristic widths $s_0$ and $\tau_0$ respectively, at a central wavelength $\lambda_0$ ($\omega_0=2\pi c/\lambda_0$). The fields of the focused RPLB with SC are modeled similarly in the frequency domain as in Ref.~\cite{sainte-marie17,jolly19-1} and the previous sections. We first must simply have the new frequency dependent coordinate $\rho'=r'/s_0=\sqrt{(x-x_0)^2+y^2}/s_0$. However a more complicated step is to apply the transformation to Cartesian coordinates together with the amplitude modifications from Eqs.~(\ref{eq:Cart1})--(\ref{eq:Cart4}) resulting in

\begin{align}
&\tilde{E}_x(\omega)=A_\textrm{SC}C_{2}\frac{\left( x-x_0 \right)}{z_R} \\
&\tilde{E}_y(\omega)=A_\textrm{SC}C_{2}\frac{y}{z_R} \\
&\tilde{E}_z(\omega)=A_\textrm{SC}\epsilon^{2}\left[S_{2}-\rho'^{2} S_{3}\right] \\
&\tilde{B}_x(\omega)=-A_\textrm{SC}C_{2}\frac{y}{c z_R} \\
&\tilde{B}_y(\omega)=A_\textrm{SC}C_{2}\frac{\left( x-x_0 \right)}{c z_R} ,
\end{align}

\noindent where $C_n$ and $S_n$ are as in Eqs.~(\ref{eq:C_n})--(\ref{eq:S_n}), with the modified $r'$ replacing $r$ in the amplitude and phase:

\begin{align}
A_\textrm{SC}&=\frac{\omega_0}{2c}\sqrt{\frac{8P_0}{\pi \epsilon_0 c}} \frac{\sqrt{2}A_\omega}{\Delta\omega} e^{-{r'}^2/s^2} \label{eq:A_SC} \\
\psi&=\Psi_0-\frac{\omega {r'}^2}{2cR}-\frac{\omega{z}}{c}-\psi_\textrm{spec} \label{eq:psi_SC}.
\end{align}

\noindent After calculating the fields in frequency space they must be inverse Fourier transformed to time as before.

\begin{figure}[tb]
	\centering
	\includegraphics[width=86mm]{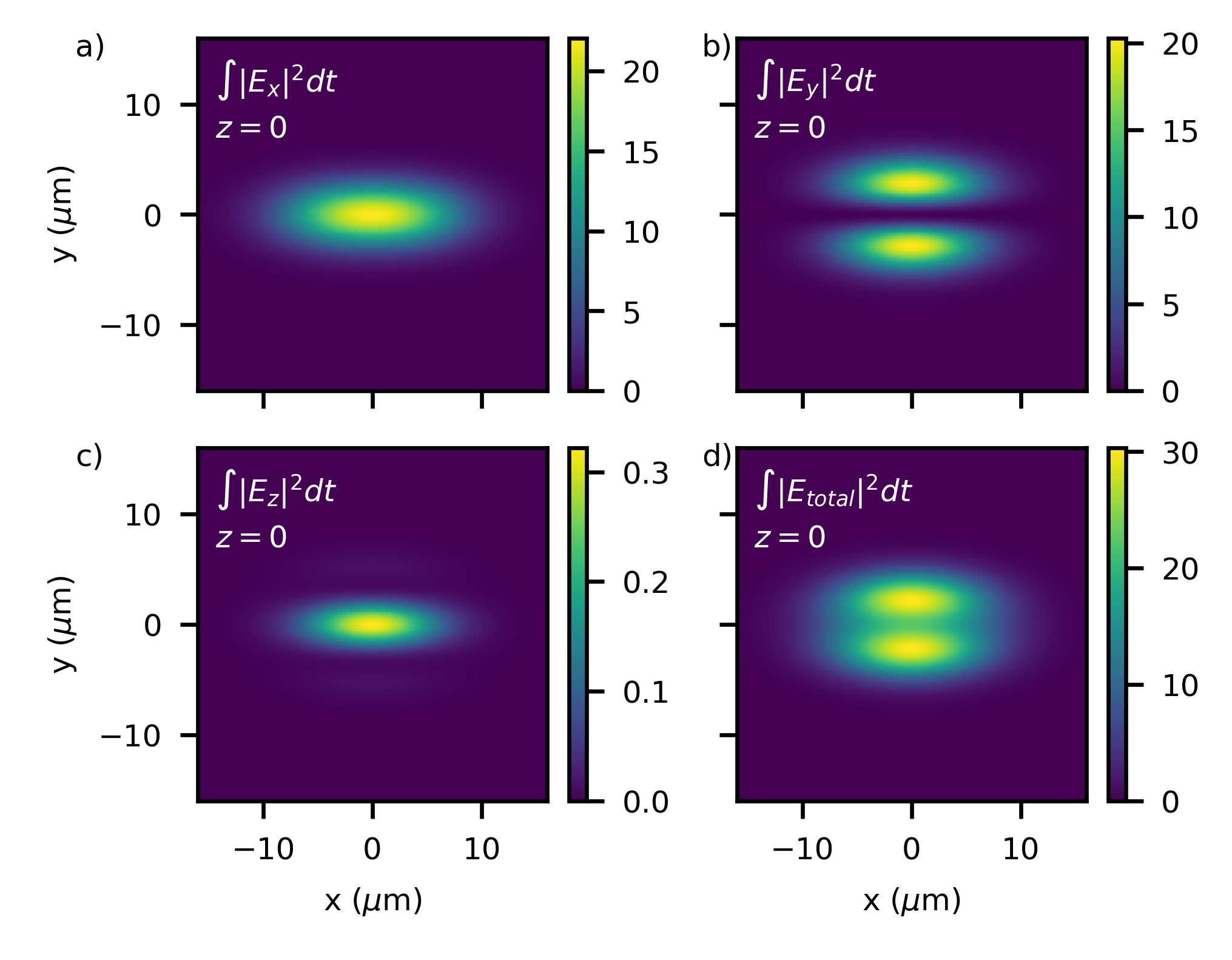}
	\caption{The intensity of the different electric field components (a) $|E_x|^2$ (b) $|E_y|^2$, and (c) $|E_z|^2$ are shown with an SC in the $x$-direction of $\tau_t=30$\,fs ($\lambda_0=800$\,nm, $\tau_0=15$\,fs, $s_0=4$\,$\mu$m). These show the asymmetry and elongation along $x$ due to the SC. The total intensity (d), which one would see on a standard camera, is also elongated along $x$ and is asymmetric. The fluence in all panels is normalized to the max fluence of $E_z$ without any STC.}
	\label{fig:SC_standard}
\end{figure}

The first most basic result is shown in Fig.~\ref{fig:SC_standard} via the fluence of the different polarization components with a single value of $\tau_t$ at the focus $z=0$. The simulated data is shown to replicate what would be seen on a camera when rejecting orthogonal polarizations in Fig.~\ref{fig:SC_standard}(a)--(b), and shows only the longitudinal component in Fig.~\ref{fig:SC_standard}(c). The total intensity shown in Fig.~\ref{fig:SC_standard}(d) replicates what would be seen on a standard camera with no analysis optics.

The qualitative results of Fig.~\ref{fig:SC_standard} agree with our expectations that with SC along $x$, $E_y$ and $E_z$ will be simply elongated along $x$. However, non-trivially, $E_x$ is extended along $x$ but also loses it's zero at $x=y=0$, seen in Fig.~\ref{fig:SC_standard}(a). Therefore not only is the total electric field in Fig.~\ref{fig:SC_standard}(d) extended along $x$, but it has more nuanced asymmetries and no longer a polarization singularity on axis at $x=y=0$.

However, beyond the simple asymmetry, there are effects on the temporal evolution of the electric field and the intensity that have potential applications and require a closer look.

\subsection{Wavefront rotation}
\label{sec:SC_WFR}

Wavefront rotation is a phenomenon with many applications with linearly-polarized laser pulses~\cite{quere14} including the attosecond lighthouse technique for generating isolated attosecond pulses~\cite{vincenti12,wheeler12,kim13,auguste16}. The wavefront rotation, defined as a changing wavefront plane in time at one point in a laser pulses propagation, is a result of SC in a focus.

Mathematically, if we follow the definition of Ref.~\cite{auguste16}, the angle of the wavefront is $\beta=(\partial\phi(x,t)/\partial{x})(c/\omega_0)$ and the rate of change of this angle, the WFR velocity $v_{r}=d\beta/dt$ [rad/s], for a linearly polarized pulse is 

\begin{equation}
v_{r}^{(\textrm{lp})}=\frac{c\tau_t}{2\omega_0 s_0}\frac{\Delta\omega^2}{1+\left(\tau_t\Delta\omega/2\right)^2},
\label{eq:WFR_linpol}
\end{equation}

with $z=y=0$ and using the notation of this manuscript.

It can be shown that this WFR velocity can be calculated directly from the fields in frequency space using the mean frequency $\bar{\omega}$ (which depends on the transverse coordinate $x$ in the case of SC) such that $v_{r}=(\partial\bar{\omega}/\partial{x})(c/\omega_0)$. This results in exactly the same relationship for RPLBs as that in Ref.~\cite{auguste16} and Eq.(\ref{eq:WFR_linpol}) for linearly polarized pulses.

Being able to calculate this WFR velocity using only frequency-space intuition is helpful for this work since we model the ultrashort RPLB field in frequency space. However, additional terms in the spectral amplitude for $E_z$ and $E_r$ mean that the WFR velocity will have a transverse dependence and the meaning of it will be more complicated due to the minima in the intensity. Rather than doing this calculation we visualize the WFR.

\begin{figure}[tb]
	\centering
	\includegraphics[width=86mm]{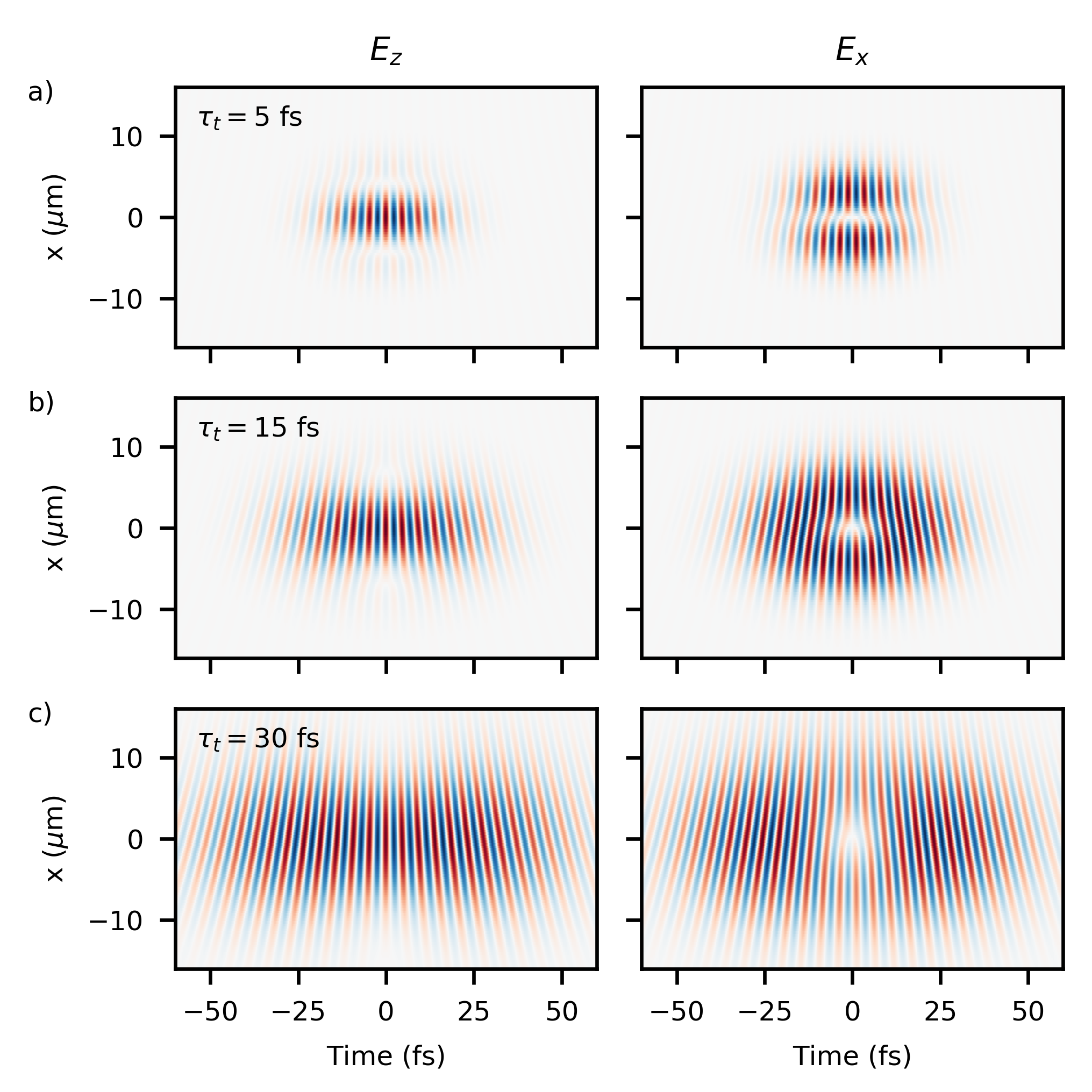}
	\caption{Varying wavefront rotation (WFR) with different spatial chirp values, parametrized by $\tau_t$, of (a) 5\,fs, (b) 15\,fs, and (c) 30\,fs ($\tau_0=15$\,fs, $s_0=4$\,$\mu$m). The WFR is clearly visible and increasing in $E_z$ (left), but is less clear in $E_x$ (right), discussed in more detail in the text.}
	\label{fig:SC_WFR}
\end{figure}

Figure~\ref{fig:SC_WFR} shows the resulting WFR in the case of $E_z$ and $E_x$ for three values of $\tau_t$. One can see complexities in both $E_z$ and $E_x$. For example, the WFR in $E_z$ dislocates briefly near the off-axis intensity minima (see left of Fig.~\ref{fig:SC_WFR}). The WFR of $E_x$ also disclocates near the axis, where there are also intensity minima (see right of Fig.~\ref{fig:SC_WFR}). Away from the axis for $E_x$ and away from the off-axis intensity minima for $E_z$ the WFR velocity is roughly uniform corresponding to Eq.(\ref{eq:WFR_linpol}). Note as well the on-axis field and amplitude is significantly modified, with a longer pulse for both $E_z$ and $E_x$ as expected, but also temporal structure in $E_x$.

\subsection{Pulse-front tilt}
\label{sec:SC_tilt}

A beam that has linear SC combined with linear temporal chirp (quadratic spectral phase or group-delay dispersion, GDD, $\phi_2$) also has pulse-front tilt~\cite{akturk04}. This is simply due to the fact that the frequencies are distributed along one spatial axis and the arrival time is also linearly dependent on frequency. This is pulse-front tilt of a different nature than AD/PFT on the collimated beam as described earlier, and since we are now looking in the focus the transverse dimension is rather $s_0$.

The central frequency of a linearly polarized pulse has the same behavior as in the previous section such that $\bar{\omega}(x)=\omega_0+x\zeta$. The temporal phase is in general not linearly dependent on the GDD $\phi_2$, but if we assume that the chirp is large, $\phi_2\Delta\omega^2\gg2$, then $\bar{\omega}(t)=\omega_0+t/\phi_2$. This results in a pulse-front tilt $\eta=dt/dx=\phi_2\zeta$ [s/m] of

\begin{equation}
	\eta=\frac{\tau_t\phi_2}{2s_0}\frac{\Delta\omega^2}{1+\left(\tau_t\Delta\omega/2\right)^2},
	\label{eq:SC_tilt}
\end{equation}

\noindent which for linearly-polarized pulses is accurate with the stated assumption of large chirp.

\begin{figure}[tb]
	\centering
	\includegraphics[width=86mm]{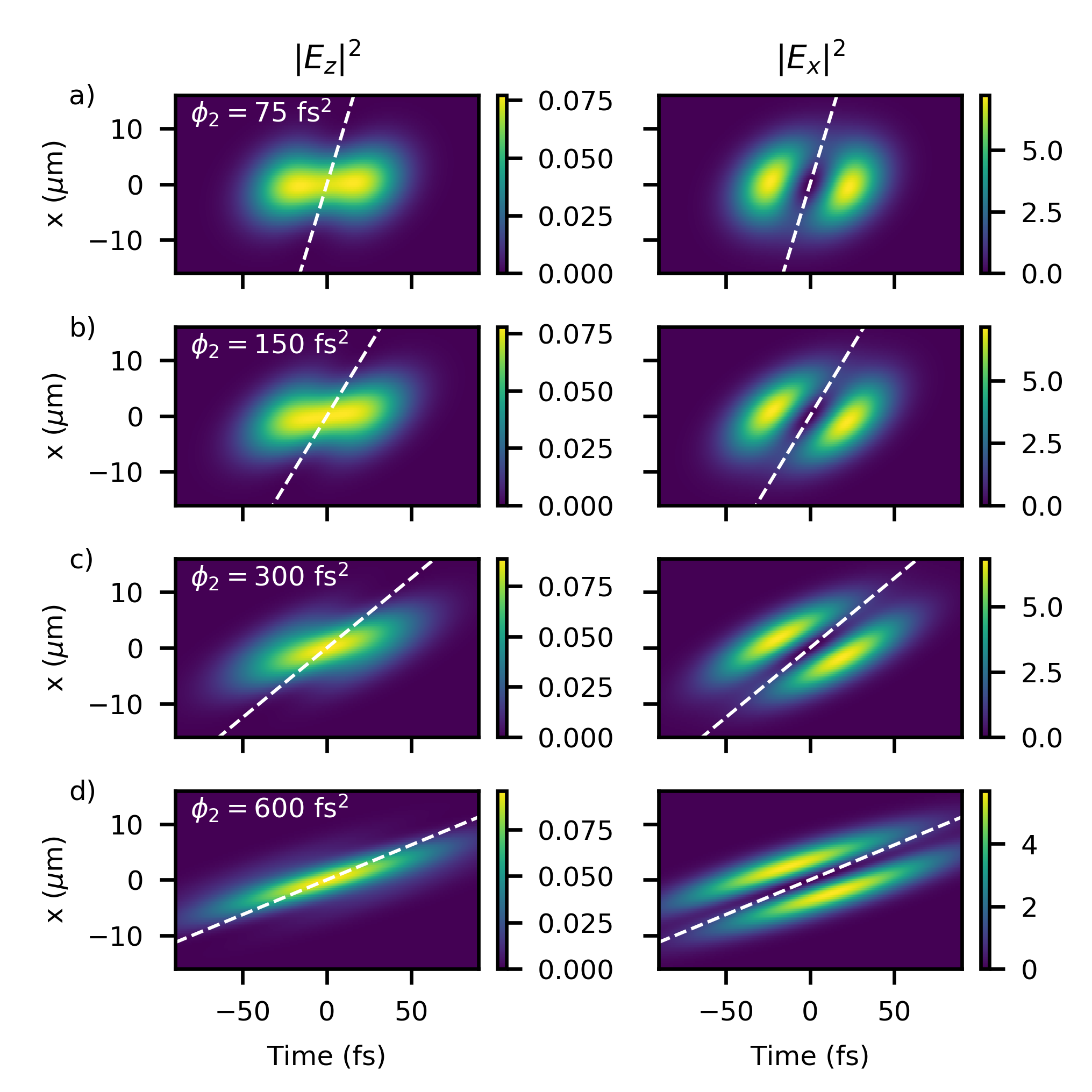}
	\caption{Adding linear chirp (nonzero $\phi_2$) to an RPLB having SC can lead to in-focus pulse-front tilt of a different nature than standard AD/PFT on a collimated beam. This is shown with $\tau_t=30$\,fs on all beams and an increasing chirp of (a) 75\,fs$^2$, (b) 150\,fs$^2$, (c) 300\,fs$^2$, and (c) 600\,fs$^2$ ($\tau_0=15$\,fs, $s_0=4$\,$\mu$m). The PFT manifests in both $|E_z|^2$ (left) and $|E_x|^2$ (right). The white dashed lines are the PFT predicted by Eq.~(\ref{eq:SC_tilt}). The intensity in all panels is normalized to the max intensity of $E_z$ without any STC.}
	\label{fig:SC_tilt}
\end{figure}

However, as was the case with WFR, the PFT is also influenced by the more complicated off-axis spectral envelope of the RPLB such that Eq.~(\ref{eq:SC_tilt}) is not fully accurate. Figure~\ref{fig:SC_tilt} shows the developing tilt at $z=0$ for a single value of $\tau_t=30$\,fs as the chirp is increased, for both $|E_z|^2$ and $|E_x|^2$. The agreement with Eq.~(\ref{eq:SC_tilt}) improves as the chirp is increased, seen clearly comparing Fig.~\ref{fig:SC_tilt}(a) and Fig.~\ref{fig:SC_tilt}(d). The agreement for $|E_z|^2$ is still not great when $\phi_2=600$\,fs$^2$, Fig.~\ref{fig:SC_tilt}(d), however for $|E_x|^2$ the agreement is very good despite the more complicated temporal intensity profile. Note counter-intuitively that the peak intensity of $E_z$ is not decreasing as chirp is added in the left column of Fig.~\ref{fig:SC_tilt}, and is in fact slightly increasing. We confirm that beyond 600\,fs$^2$ the peak intensity does begin to decrease for the specific parameters of Fig.~\ref{fig:SC_tilt}, and that for larger values of $\tau_t$ the peak intensity is maximum at a larger chirp. This interesting result potentially allows for designed aberrations with a partially mitigated intensity decrease.

\section{Frequency-dependent beam parameters}
\label{sec:g0}

In the previous two models the beam width and Rayleigh range have been approximated as constant values related to the central frequency and the pulse duration was such that the RPLB with no STCs had many cycles. However, diffraction itself and especially tight focusing is a chromatic phenomenon, and therefore the waist and Rayleigh range are in reality frequency dependent properties: $z_R={\omega}s_0^2/2c$ is a fixed relationship where either $z_R$ or $s_0$ can have frequency dependence according to the specifics of the physical situation. This phenomenon was known early on to effect the reshaping of single-cycle pulses through the focus~\cite{christov85,ziolkowski92,feng98,porras02}. The detailed effect of a non-uniform spatio-spectral beam width on the phase in the focus of ultrashort pulses---the so-called focal-phase---has been explored in-depth with linearly-polarized pulses~\cite{porras09,porras12,hoff17-1,porras18} and shown to have an effect on photoelectron production driven by such ultrashort laser pulses~\cite{hoff17-2,zhangY20}. A model with RPLBs, similar to that presented here, was used to show that the effect is significant as well in vacuum laser acceleration~\cite{jolly20-2}.

The basics of the model are first that Eqs.~(\ref{eq:E_r})--(\ref{eq:R}) for the standard RPLB are modified with frequency-dependent Rayleigh range $z_R(\omega)$ and waist $s_0(\omega)$. In order to encapsulate the frequency dependence we use the "Porras factor" $g_0$:

\begin{equation}
g_0 = -\frac{dz_R(\omega)}{d\omega}\biggr\rvert_{\omega_0} \frac{\omega_0}{z_R(\omega_0)}.
\end{equation}

\noindent We use the reference parameter of the beam waist at the central frequency $s_0(\omega_0)=s_{00}$, which results in the equations for the beam parameters

\begin{align}
z_R(\omega)=z_{R0}\left(\frac{\omega_0}{\omega}\right)^{g_0}=\frac{\omega_0 s_{00}^2}{2c}\left(\frac{\omega_0}{\omega}\right)^{g_0} \label{eq:z_R} \\
s_0(\omega)=s_{00}\left(\frac{\omega_0}{\omega}\right)^{\frac{g_0+1}{2}} \label{eq:w_0}.
\end{align}

\begin{figure}[tb]
	\centering
	\includegraphics[width=86mm]{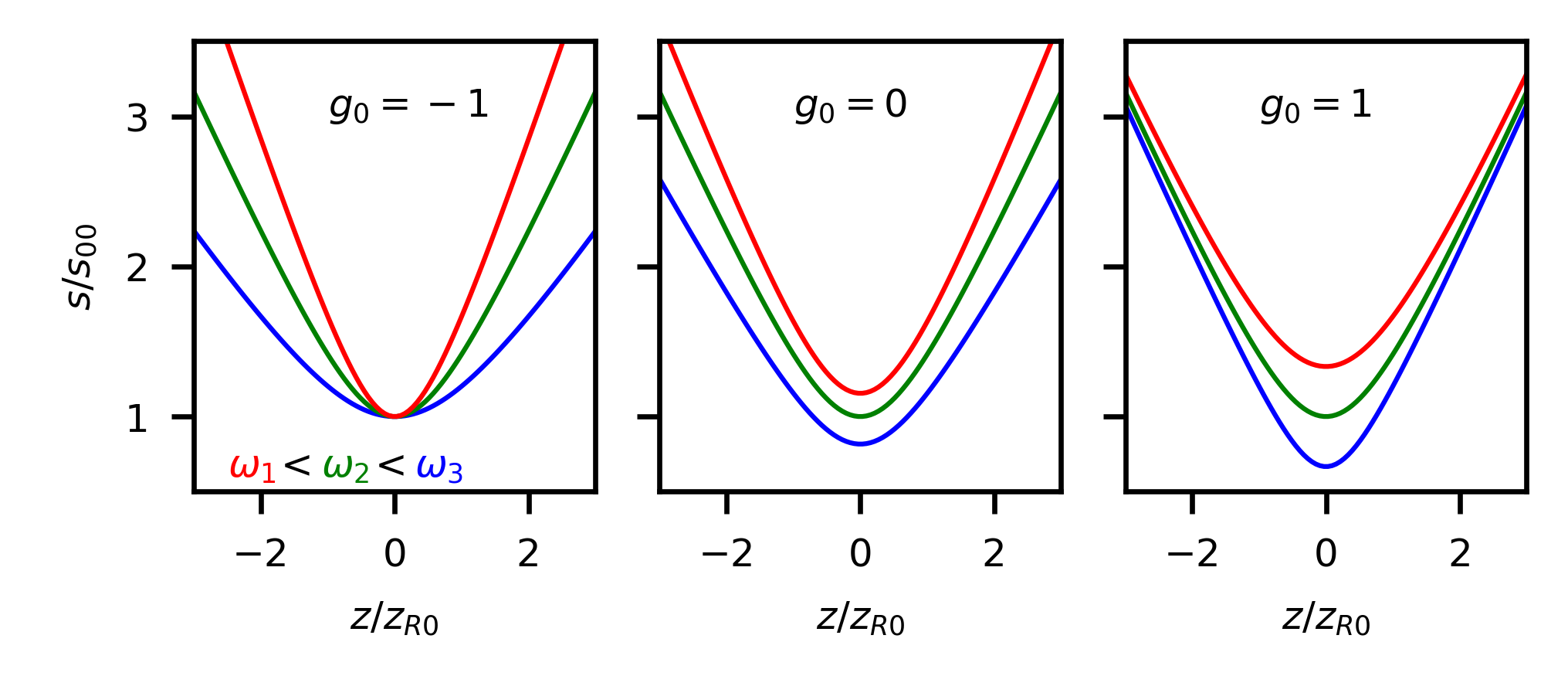}
	\caption{The basic concept of $g_0$ shown with three frequencies and three $g_0$ values. When $g_0=-1$ (left) all frequencies have the same beam size in the focus, where in contrast when $g_0=1$ (right) all three frequencies diverge with the same opening angle outside of the focus. $g_0=0$ (middle) is an intermediate situation.}
	\label{fig:g0_concept}
\end{figure}

\noindent This is shown for three frequencies in the case of $g_0=$-1, 0, and 1 in Fig.~\ref{fig:g0_concept}. 

In order for the effects of the frequency-dependent parameters to be significant, the temporal duration must approach just a few optical cycles. In this regime the Gaussian spectral envelope is no longer strictly valid. We use a Poisson-like spectrum as follows

\begin{equation}
A_{\omega}(\omega) = \sqrt{\frac{2}{\pi}}\left(\frac{\gamma\omega}{\omega_0}\right)^{\gamma+1}\left(\frac{1}{\omega}\right)\frac{e^{-\gamma\omega/\omega_0}}{\Gamma(\gamma+1)} \label{eq:Poisson},
\end{equation}

\noindent where $\gamma$ is the parameter that now represents the pulse duration, with $\tau_0=\gamma\sqrt{e^{(2/(\gamma+1))}-1}/\omega_0$ ($1/e^2$ intensity width)~\cite{caron99}, and $\Gamma(...)$ is the Gamma function for the purely real argument $\gamma+1$.

Using these frequency-dependent parameters (Eqs.~(\ref{eq:z_R})--(\ref{eq:w_0})) and Poisson-like spectral envelope (Eq.~(\ref{eq:Poisson})) in place of the relevant terms in Eqs.~(\ref{eq:E_r})--(\ref{eq:R}), and Fourier transforming to time, produces the fields without the common frequency-independent approximation.

The main result of properly modeling the frequency-dependent beam parameters is that the carrier-envelope phase (CEP) varies through the focus in a way non-trivially different than the standard Guoy phase. The key difference with the RPLB is that this behavior is not the same for the different polarization components. The CEP evolution $\Delta\Psi$ through the focus is calculated using the relation $\Delta\Psi=\phi\rvert_{\omega_0}-\omega_0\frac{\partial\phi}{\partial\omega}\rvert_{\omega_0}$ (the difference between the pulse-front and the wavefront) where $\phi$ is the total phase such that

\begin{align}
\Delta\Psi_r(r,z) &= \frac{g_0\left[2-2\frac{r^2}{s^2}\right]}{\frac{z}{z_{R0}}+\frac{z_{R0}}{z}} - 2\tan^{-1}\left(\frac{z}{z_{R0}}\right) \label{eq:g0_CEP_r} \\
\Delta\Psi_z(r=0,z) &= \Delta\Psi_r(r=0,z) \label{eq:g0_CEP_z0}.
\end{align}

\noindent Note that $s$ is the usual frequency-independent value as in Eq.~\ref{eq:w} and the last term in Eq.~(\ref{eq:g0_CEP_r}) is simply double the frequency-independent Gouy phase as in Eq.~\ref{eq:psi_G}. Due to the complexity of the phase for $E_z(r{\neq}0,z)$, the expression for the CEP evolution is not solved for explicitly in that case except for when $r=0$, where it is in fact equal to that of $E_r$.

Besides the CEP evolution through the focus, the central frequency also evolves. This is easy to visualize since, as depicted in Fig.~\ref{fig:g0_concept}, the different colors are focused to different waists depending on the $g_0$ value. Regardless of the assumed integrated spectral envelope $A_{\omega}$, the central frequency must evolve through the focus since the colors that are larger outside of the focus are the smallest within the focus and vice-versa.

\begin{figure}[tb]
	\centering
	\includegraphics[width=86mm]{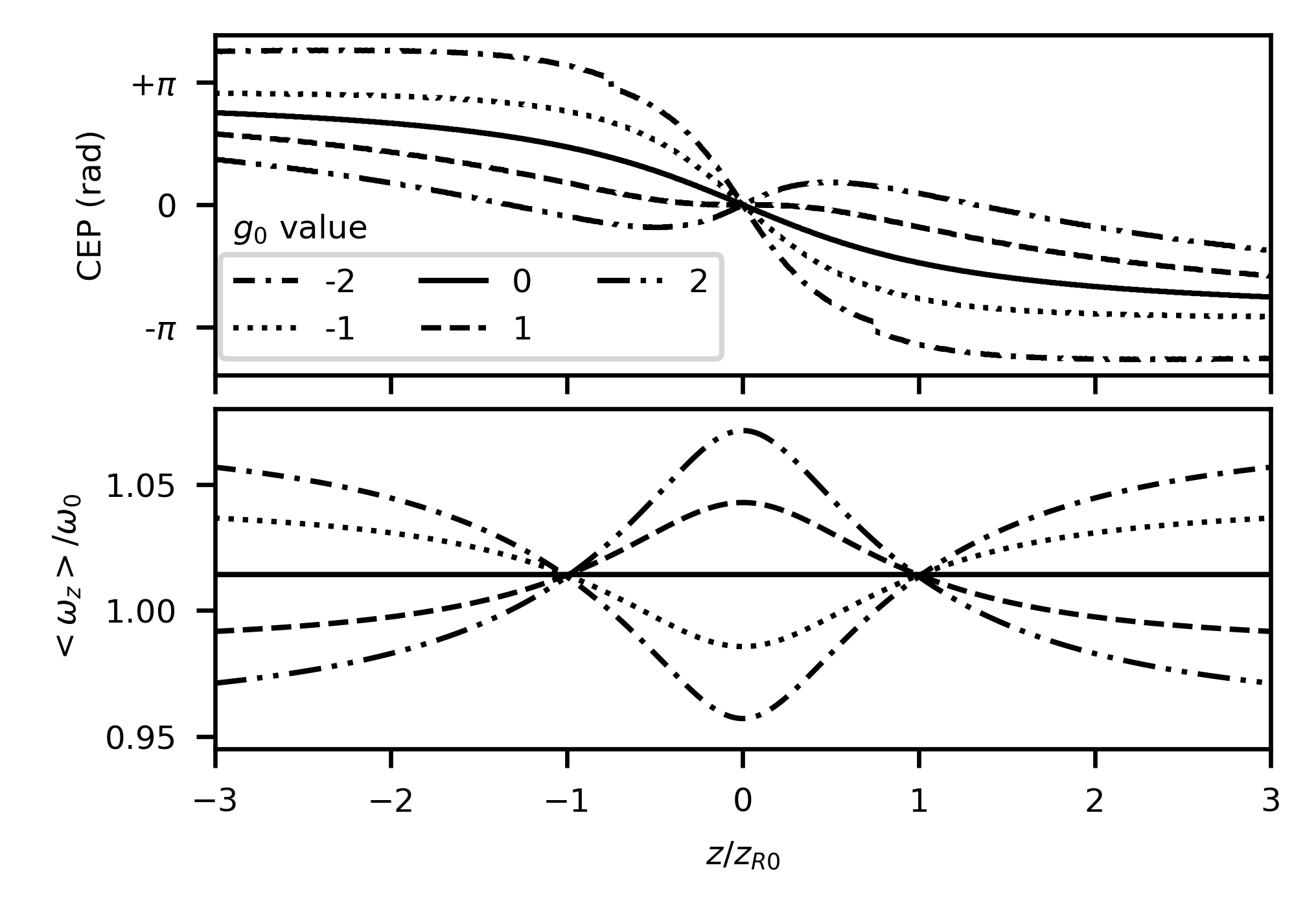}
	\caption{On-axis CEP and central frequency evolution through the focus with a duration of 3.5\,fs ($\gamma$=35). The CEP (top) varies widely, showing a steep slope at $g_0=-2$ and an inflection when $g_0\geq{1}$. The central frequency (bottom) also varies through the focus due to the reshaping of the frequency envelope as different frequencies are focused more tightly. This figure is identical to one presented in ref.~\cite{jolly20-2}.}
	\label{fig:g0_CEP}
\end{figure}

The on-axis ($r=0$) CEP evolution, identical for $E_r$ and $E_z$ as in Eq.~(\ref{eq:g0_CEP_z0}), is shown in Fig.~\ref{fig:g0_CEP}(top), and the central frequency evolution for $E_z$ is shown in Fig.~\ref{fig:g0_CEP}(bottom) for $\tau_0=3.5$\,fs ($\gamma$=35). It is important to note that the CEP evolution does not change with the pulse duration, it only has a greater significance as the pulse duration decreases, but the central frequency evolution does depend strongly on the pulse duration~\cite{porras09}.

\begin{figure}[tb]
	\centering
	\includegraphics[width=86mm]{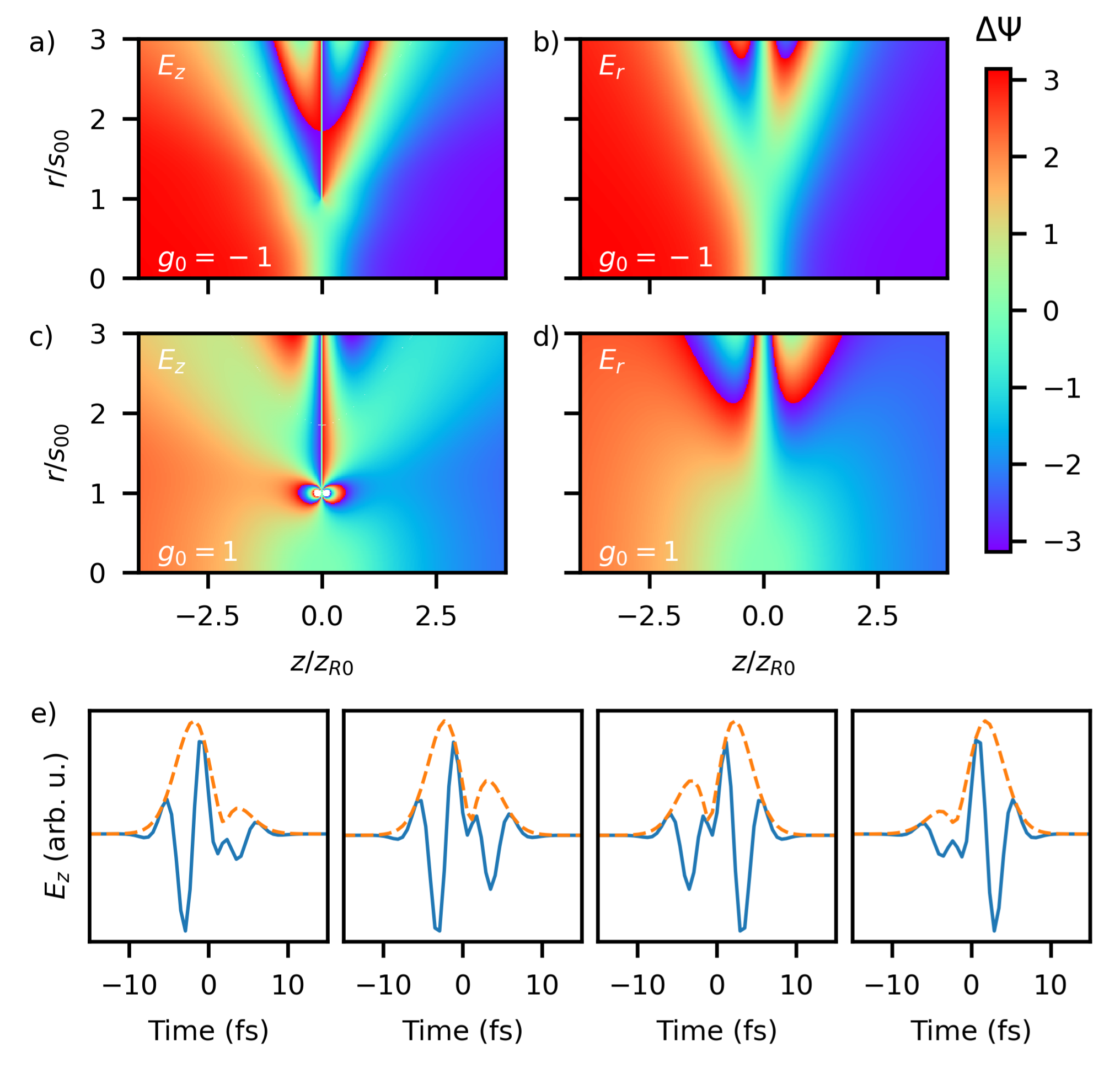}
	\caption{Off-axis CEP evolution through the focus. The CEP for both $E_z$ ((a) and (c)) and $E_z$ ((b) and (d)) are shown through the focus for on- and off-axis positions, with a $g_0$ of -1 ((a) and (b)) and +1 ((c) and (d)). The discontinuities in the CEP of $E_z$ are due to pulse reshaping around $r/s_{00}=1$ and $z=0$, shown in (e) for a pulse duration of 3.5\,fs ($\gamma$=35) with $g_0=1$ at four $z/z_{R0}$ positions of -1/4, -1/8, +1/8, +1/4 from left to right, with the field (blue solid line) and the amplitude (orange dashed line).}
	\label{fig:g0_CEP_2D}
\end{figure}

These on-axis results in Fig.~\ref{fig:g0_CEP} were already reported in the context of the application to vacuum acceleration of electrons~\cite{jolly20-2}. However, the off-axis CEP evolution has not yet been reported and is shown here in Fig.~\ref{fig:g0_CEP_2D} for $E_z$ and $E_r$ for both $g_0=-1$ and +1. The plotted results for $E_r$ are simply that of Eq.~(\ref{eq:g0_CEP_r}), while the results for $E_z$ are calculated via the numerical differentiation of the phase of the field.

The first result to note is that the difference between the phase for varying $g_0$ clearly extends to off-axis positions as well both for $E_z$ and $E_r$, most clear when comparing off-axis positions in Fig.~\ref{fig:g0_CEP_2D}(b) to Fig.~\ref{fig:g0_CEP_2D}(d). The second notable result is that for $E_z$ there are significant discontinuities as $r$ increases. Specifically at $r/s_{00}=1$ and $z=0$ the discontinuity is due to local reshaping of the field as shown in Fig.~\ref{fig:g0_CEP_2D}(e), which exposes a weakness of the simple method to calculate the CEP change. One could extend the analysis away from the implicit non-reshaping assumption as in Ref.~\cite{porras12}, but in this specific region $E_z$ becomes very small and is therefore less important, and due to the very significant reshaping the meaning of the CEP is less clear. We therefore leave a more nuanced analysis to a further work.

\section{Conclusion}
\label{sec:conclusion}

We have presented the fields of ultrashort radially-polarized laser pulses when tightly focused and having three different low-order spatio-temporal couplings: longitudinal chromatism, spatial chirp, and frequency-dependent beam parameters. We presented results mostly directly in the focus but also included some propagation characteristics. This was presented for all three cases using a model where the effect of the STCs was developed by adding frequency-dependence to parameters of the description in frequency-space and confirming the existence of important effects in the time domain via numerical Fourier transformation. The lack of an analytical description directly in the time domain is an important aspect of ongoing work, and we hope that drawing upon the descriptions of other fundamental ultrashort pulses that have interesting space-time properties such as tilted beams~\cite{wong17-2}, the flying donut~\cite{ziolkowski89,hellwarth96,zgadkas20}, or the previously mentioned space-time light sheets~\cite{kondakci17} may address this issue. Still, we believe that developing this model and recipe for describing low-order STCs and showing the powerful effects that take place in the focus of ultrashort radially-polarized laser beams is useful without the time domain equations. 

Further complexities can be imagined beyond what was presented in this manuscript. The first possibility of course is including higher-order STCs, such as LC or SC that are nonlinearly dependent on the frequency or frequency-dependent beam width that cannot be described by one parameter, or indeed a simple combination of the presented STCs. Or, for example, combining the presented STCs with more realistic near-field profiles for high-power laser beams such as the super-Gaussian or flattened-Gaussian~\cite{bagini96}, or modelling higher-order vector beams, or using the Richards–Wolf formalism~\cite{wolf59,richards59} to model very non-paraxial beams or beams diffracting from an aperture, important for the Guoy phase of RPLBs~\cite{pelchat-voyer20}. Additionally, there may be nuanced behavior of either the phase or the electric field when including a more detailed analysis of the RPLBs outside of the focus. The models presented here are general and provide the framework for investigating these more complex situations.

We believe that the combination of vector beams and STCs will become of increasing interest for ultrafast optical applications as the knowledge and understanding becomes more advanced. The combination of the models and analysis presented in this work and the further development of methods to control and characterize STCs~\cite{jolly20-3}---also for vector beams~\cite{alonso20}---will enable this advancement.

\section*{Acknowledgements}

Example scripts to generate the fields presented in this work are available in a public repository~\cite{github_RPLB-STC}.

The author would like to acknowledge Fabien Qu{\'e}r{\'e} for helpful discussions and Martin Virte for input on the manuscript.

\section*{Appendix: Non-paraxial description}

Here we outline the description for non-paraxial beams according to the method used in Ref.~\cite{salamin06} to add to the paraxial equations and results presented in the main text. The chosen method, without STCs, expands the wave equation to higher orders of the generally small parameter $\epsilon=s_0/z_R=\lambda_0/\pi{s_0}$ to have more accurate descriptions of the fields, improving upon Eqs.~\ref{eq:E_r}--\ref{eq:B_t}, with all of the same properties unchanged from Eqs.~\ref{eq:C_n}--\ref{eq:R}. Specifically,

\begin{align}
\begin{split}
&\tilde{E}_r=A_\textrm{np}\Bigg\{\epsilon\left[\rho C_{2}\right]+\epsilon^{3}\left[-\frac{\rho C_{3}}{2}+\rho^{3} C_{4}-\frac{\rho^{5} C_{5}}{4}\right] \\
&+\epsilon^{5}\left[-\frac{3 \rho C_{4}}{8}-\frac{3 \rho^{3} C_{5}}{8}+\frac{17 \rho^{5} C_{6}}{16}-\frac{3 \rho^{7} C_{7}}{8}+\frac{\rho^{9} C_{8}}{32}\right]\Bigg\}
\end{split}\\
\begin{split}
&\tilde{E}_z=A_\textrm{np}\Bigg\{\epsilon^{2}\left[S_{2}-\rho^{2} S_{3}\right] \\
&+\epsilon^{4}\left[\frac{S_{3}}{2}+\frac{\rho^{2} S_{4}}{2}-\frac{5 \rho^{4} S_{5}}{4}+\frac{\rho^{6} S_{6}}{4} \right] \Bigg\}
\end{split}\\
\begin{split}
&\tilde{B}_{\theta}=\frac{A_\textrm{np}}{c}\Bigg\{\epsilon\left[\rho C_{2}\right]+\epsilon^{3}\left[\frac{\rho C_{3}}{2}+\frac{\rho^{3} C_{4}}{2}-\frac{\rho^{5} C_{5}}{4}\right] \\
&+\epsilon^{5}\left[\frac{3 \rho C_{4}}{8}+\frac{3 p^{3} C_{5}}{8}+\frac{3 \rho^{5} C_{6}}{16}-\frac{\rho^{7} C_{7}}{4}+\frac{\rho^{9} C_{8}}{32}\right] \Bigg\}.
\end{split}
\end{align}

\noindent The only additional correction is on the beam power, of the form

\begin{align}
A_\textrm{np}&=\frac{\omega_0}{2c}\sqrt{\frac{8P_\textrm{np}}{\pi \epsilon_0 c}} \frac{\sqrt{2}A_\omega}{\Delta\omega} e^{-{r}^2/s^2} \label{eq:A_np} \\
P_\textrm{np}&=\frac{P_0}{1+3\left(\frac{\epsilon}{2}\right)^2+9\left(\frac{\epsilon}{2}\right)^4} \label{eq:P_np},
\end{align}

\noindent where $P_0$ is the physical power of the beam, equal to the power in the non-paraxial case (without STCs). It must be noted that while this description is more accurate in describing the fields for moderate values of $\epsilon$, the convergence is not well understood as $\epsilon$ becomes comparable to 1 since it is based on an expansion in $\epsilon$ as a small parameter.

When longitudinal chromatism (LC) is present $z$ is replaced everywhere with $z-z_0(\omega)$ which effects the same terms $\psi_G$, $s$, and $R$ as in Eqs.~\ref{eq:psi_G_LC}--\ref{eq:R_LC}. This results in both the amplitude and phase of all $C_n$ and $S_n$ being frequency dependent, with the only difference being that in the non-paraxial case there are many more terms of higher $n$.

In the case where spatial chirp (SC) is present the modifications are also as before in the paraxial case, effecting the amplitude of all of the fields as in Eq.~\ref{eq:A_SC} and the phase as in Eq.\ref{eq:psi_SC} due to the modified $r'$, and also requiring a frequency-dependent coordinate transformation as in Eqs.~\ref{eq:Cart1}--\ref{eq:Cart4}. The main difference then for the non-paraxial case is the higher orders of $\rho$ that now need to be described by the frequency-varying $\rho'$, and as for LC the phase effects are present as well in the higher orders of $C_n$ and $S_n$.

For the third and final STC described in the main text, frequency-varying beam parameters, $z_R$ and $s_0$ need to be replaced by their frequency-dependent form depending on the $g_0$ value as in Eqs.~\ref{eq:z_R}--\ref{eq:w_0}. This introduces frequency dependence in $\epsilon$, $\rho$, $s$, $\psi_G$, and $R$ which in sum effect both the phase and amplitude. And again, the primary additional complexity in the non-paraxial case is the higher orders present of $\epsilon$, $\rho$, $C_n$, and $S_n$.

In summary, the non-paraxial equations have been presented in this appendix and a recipe for adapting each of the three STCs presented in the main text to the non-paraxial form has been given here. However, this could only be in a general matter since in all cases both the amplitude and phase develop frequency dependence that is of much higher order in the non-paraxial form and therefore non-trivial. It must be noted that for the parameters presented in the main text ($\lambda_0$=800\,nm, $s_0$=4\,$\mu$m, $\epsilon$=0.0637) the non-paraxial results are qualitatively the same, confirmed by numerical calculations, and the fields and intensity are very similar quantitatively. Only for extremely sensitive interactions such as vacuum electron acceleration would the non-paraxial fields be important with such focusing~\cite{marceau13-1}. Indeed for tighter focusing (larger $\epsilon$) the non-paraxial form becomes more important, and would begin to significantly effect especially the field-based effects presented in the main text, specifically wavefront rotation and the changing CEP through the focal volume.

\bibliographystyle{unsrt}
\bibliography{biblo}

\end{document}